# One-step method to grow $Ba_{0.6}K_{0.4}Fe_2As_2$ single crystals without fluxing agent

**Chunlei Wang, Chao Yao, Lei Wang, Yanpeng Qi, Zhaoshun Gao, Dongliang Wang, Xianping Zhang and Yanwei Ma[1]**

Key Laboratory of Applied Superconductivity, Institute of Electrical Engineering, Chinese Academy of Sciences, PO box 2703, Beijing 100190, China

## Abstract

Single crystals of $Ba_{0.6}K_{0.4}Fe_2As_2$ with excellent quality have been successfully grown without fluxing agent through a simple one-step method for the first time. X-ray diffraction patterns demonstrate that the samples have high crystalline quality and *c*-axis orientation. The onset transition temperature is up to 38 *K* with the zero resistivity temperature about 36.7 *K*. Both the *R-T* and *M-T* data show a very sharp superconducting transition with transition width 0.4 *K*. We also found that the samples possess very large current carrying ability and high upper critical fields, indicating potential applications requiring very high field. The above simple and safe one-step technique of single crystal growth can be effective in other systems of Fe-based superconductors.

[1] Author to whom any correspondence should be addressed.
E-mail: ywma@mail.iee.ac.cn

## Introduction

The newly discovered superconductivity in $LaFeAsO_{1-x}F_x$ with $T_c$ about 26 $K$ has stimulated intensive research on high temperature superconductivity besides the cuprate compounds [1]. To today, various Fe-based superconductors have been found with the highest $T_c$ about 55 $K$ [2-5], marking the highest record outside the cuprate system. Among these novel Fe-based superconductors, the $Ba_{1-x}K_xFe_2As_2$ has attracted much more significant coverage from scientists because of its relative high superconducting critical temperature (38 $K$) and low sintering-temperature [6]. Though $BaFe_2As_2$, like LaFeAsO, is a tetragonal compound with a Fe plane capped above and below As, it does not contain oxygen. Thus, compared to these '1111' superconductors, it is much more convenient to prepare single crystal samples. For example, references [7-8] have reported that single crystals with *mm* size could be obtained by Sn flux method. However, the $Ba_{1-x}K_xFe_2As_2$ single crystals obtained by this method usually contain some Sn impurities [7]. In order to avoid the contamination of Sn, Wang, et al [9] has produced sizable single crystals of $BaFe_2As_2$ with self-flux method. Luo, et al [10-11] also prepared high quality $Ba_{1-x}K_xFe_2As_2$ single crystals using this method. However, there are always a lot of FeAs impurities between the $Ba_{1-x}K_xFe_2As_2$ superconducting phases fabricated by this method, which will confine its applications.

Recently, Zhang, *et al* has successfully grown single crystal of $BaFe_{2-x}Co_xAs_2$ with $T_c$ about 23 $K$ without fluxing agent through a three-step method [12]. Compared to $BaFe_{2-x}Co_xAs_2$ single crystals, $Ba_{1-x}K_xFeAs$ has higher superconducting transition temperature up to 38 $K$ [13,14]. Thus, it is much more significant to develop a simple and effective method to fabricate $Ba_{1-x}K_xFeAs$ single crystals for both theory and applications. However, the growth of $Ba_{1-x}K_xFeAs$ single crystals usually faces much more challenges than that of $BaFe_{2-x}Co_xAs_2$, for instance, the silica tube would break due to the vapour pressure of potassium at high temperature, and potassium can easily react with silica tube [10]. All these factors, together with burning loss at high temperature, make it very difficulty to control the content of potassium. Here we report a safe and simple one-step method to grow $Ba_{1-x}K_xFe_2As_2$ single crystal without fluxing agent. The superconducting transition is sharp which points to optimistic superconducting properties.

## Experimental details

The single crystals of $Ba_{0.6}K_{0.4}Fe_2As_2$ were synthesized through a simple one-step method. Stoichiometric amount of Ba filings, Fe powder, As and K pieces, were ground in Ar atmosphere for more than 6 hours, with the aim to achieve a uniform distribution. In order to compensate the evaporation of As and K at high temperatures, extra As and K were added. To avoid the reaction between the K vapor and quartz glass, we have employed an improved method using a commercial stainless steel container as displayed in Fig. 1. Then an alumina crucible containing the starting materials was put into the steel tube. Tight sealing was accomplished by two caps also made of stainless steel which were welded to the both ends of the tube by argon-arc welding. Then the samples were placed in a tube furnace and heated up to a high temperature (above 1180$^{o}$C) to melt completely. Then samples were cooled down to temperature below 1050$^{o}$C. Finally, the furnace was powered off. In order to reduce oxidation of the samples, high purity argon gas was allowed to flow into the furnace during the sintering processes. The phase identification was characterized using x-ray diffraction (XRD) analysis with Cu-Kα radiation from 20 to 80$^{o}$. The diffraction peaks could be well indexed on the basis of tetragonal $ThCr_2Si_2$-type structure with the space group I4/mmm. EDX analysis system was employed to determine the content of the single crystals. The superconducting properties were studied by magnetization and standard four-probe resistivity measurements using a physical property measurement system (PPMS). The critical current density $J_c$ was determined using the Bean model.

## Results and discussion

Fig. 2 shows a photograph of some crystals cleaved from the as-grown bulks. Clearly, all samples have very shiny plate-like cleaved surfaces. The size of the largest single crystal is about 0.5 mm × 4 mm × 6 mm, as shown in Fig. 2. It should be noted that the real sizes of the crystals are obviously larger than that shown in Fig. 2, but it is much harder to cleave the crystals grown by the present method than that grown by self-flux method. Several pieces of as-grown single crystals are selected. Then they are cleaved and EDX analyses were preformed, as shown in Fig. 3. The actual atomic ratios are very close to stoichiometric compound of $Ba_{0.6}K_{0.4}Fe_2As_2$. The inset is the SEM photograph of the crystal showing a very flat surface morphology.

In order to judge the quality of the samples, X-ray diffraction (XRD) measurement is employed. Fig. 4 presents the typical XRD patterns of $Ba_{0.6}K_{0.4}Fe_2As_2$ single crystals. Only the (00*l*) peaks with even *l* are observed, suggesting that the crystallographic *c*-axis is perpendicular to the shining surface. The sharp diffraction peaks indicate good quality of single crystal.

Figure 5 shows the temperature dependence of resistivity for a $Ba_{0.6}K_{0.4}Fe_2As_2$ single crystal in zero field. As we can see, resistivity drops at 38.2 *K* and vanishes at about 36.7 *K*. The resistivity data also indicate a sharp transition for our samples with transition width about 0.4 *K* (90% - 10% of normal state resistivity). The residual resistivity ratio (*R(*300*K)/R(*39*K)*) is large up to ~8. In addition, if we extrapolate the data just above the superconducting transition through a straight line, it would be roughly estimated that residual resistivity is almost close to zero for our samples, indicating that our samples are rather clean. The inset of Fig. 5 exhibits the typical susceptibility curves. There is a flat diamagnetic signal in the low-temperature region and there is also a very sharp superconducting transition around $T_c$. It is also found that the bulk magnetic susceptibility above $T_c$ is nearly temperature-independent, meaning that the paramagnetic state is a Pauli-paramagnetic state, which is consistent with the metallic behavior of $Ba_{0.6}K_{0.4}Fe_2As_2$ system. Both the *R-T* and *M-T* experimental results confirm that our samples have excellent quality.

We also measure the temperature dependence of resistivity at various magnetic fields for *H* parallel to *c*-axis, and the results are shown in Fig. 6. It can be seen that $H_{c2}$ and $H_{irr}$ values for $Ba_{0.6}K_{0.4}Fe_2As_2$ are quite high. For example, the $H_{c2}$ at 34.9 *K* and $H_{irr}$ at 33.9 *K* are 9 T, which are slightly higher than that reported by references [13-15]. $H_{c2}(0)$ is estimated by using the Werthamer-Helfand-Hohenberg (WHH) formula: $H_{c2}(0) = -0.693T_c \, {dH_{c2}}/{dT}$, with ${dH_{c2}}/{dT}$ at $T = T_c$. Using the criterion of the upper critical field $H_{c2}$, which is taken from the resistive transition by 90% $R_n$, where $R_n$ denotes the normal state resistance. The rough estimated $H_{c2}(0)$ is more than 140 *T* for our samples in applied field parallel to *c*-axis. If we take $R = 10\%R_n$ as the criterion of the irreversibility field, then the relationship between irreversibility field and temperature is also displayed, as shown in inset of Fig. 6. It is clear the that $H_{irr}(0)$ should be slightly smaller than $H_{c2}(0)$.

Fig. 7 shows the magnetic field dependence of the critical current density $J_c$ for a $Ba_{0.6}K_{0.4}Fe_2As_2$ single crystal, derived from the hysteresis loop width using the extended Bean model of $J_c = 20\Delta M / Va(1 - a/3b)$ taking the full sample dimensions. Where $\Delta M$ is the height of magnetization loop, *a* and *b* denote the dimensions of the sample perpendicular to the direction of magnetic field ($a < b$). Note that the calculated *J*c at 5 *K* and 20 *K* gets up to about $9\times10^5$ A/cm$^2$ and $6.5\times10^4$ A/cm$^2$ at applied field 8 T, respectively, indicating a potential applications in industries.

## Conclusion

In summary, we have grown large-size $Ba_{0.6}K_{0.4}Fe_2As_2$ single crystals without using any fluxing agent by simple one-step method. The samples have sizes up to 6 mm with flat and shining cleaved surfaces. The X-ray diffraction patterns with only (00*l*) peaks suggesting high crystalline quality. The magnetic susceptibility and resistivity measurements display very sharp superconducting transition, indicating excellent superconducting properties. The large critical current density and high upper critical field exhibit a fascinating application prospect.

## Acknowledgments

The authors thank Haihu Wen, Liye Xiao and Liangzhen Lin for their help and useful discussion. This work is partially supported by the Beijing Municipal Science and Technology Commission under Grant No. Z07000300700703, National Science Foundation of China (Grant No. 50802093) and National ‘973’ Program (Grant No. 2011CBA00105).

## Figures

Fig. 1 Sealing assembly for the single crystal preparation.(1) Stainless tube, (2) stainless caps, (3) alumina crucible, and (4) Raw materials

Fig. 2 Photograph of the $Ba_{0.6}K_{0.4}Fe_2As_2$ single crystals cleaved from the as-grown bulk. The crystals have rather shiny surfaces with sizes up to 6 mm.

Fig. 3 A typical EDX spectrum for one single crystal. The inset is the SEM photograph of this crystal, which shows a very flat surface morphology.

Fig. 4 Typical XRD patterns for cleaved crystals.

Fig.5 Temperature dependence of resistivity for $Ba_{0.6}K_{0.4}Fe_2As_2$ single crystal. $\delta T_c$ is defined as $T(R^N_{90\%}) - T(R^N_{10\%})$. $R^N_{90\%}$ and $R^N_{10\%}$ indicate the 90% - 10% of normal state resistivity, respectively. The inset shows Temperature dependence of the magnetic susceptibility for the $Ba_{0.6}K_{0.4}Fe_2As_2$ single crystal. Measurement is done under the magnetic field of 20 Oe applied parallel to the *c*-axis.

Fig. 6 Temperature dependence of resistivity at various fields of 0, 1, 3, 5, 7, and 9 T of $B_{a0.6}K_{0.4}Fe_2As_2$ single crystal for *H* parallel to *c*-axis. Inset: The upper critical field $H_{c2}$ and irreversibility field $H_{irr}$ as a function of temperature.

Fig. 7 Magnetic field dependences of critical current densities at 5 K and 20 K for $B_{a0.6}K_{0.4}Fe_2As_2$ single crystal.

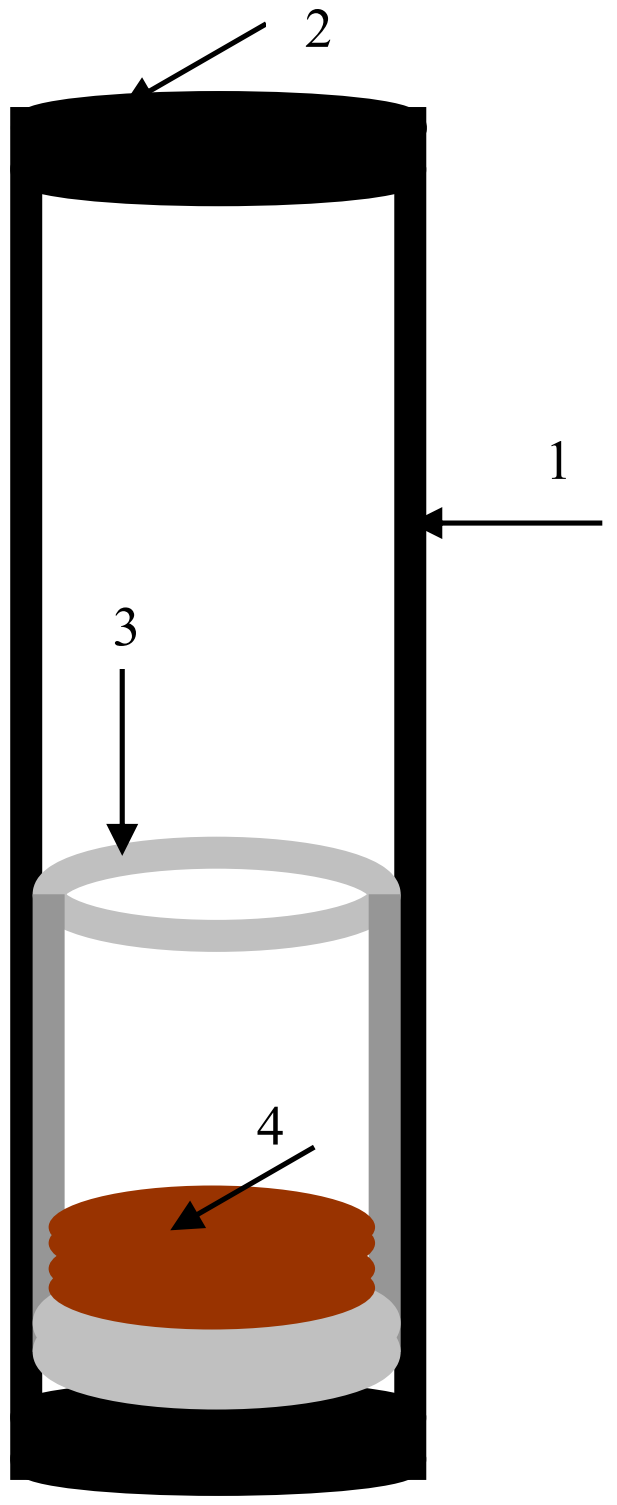

Fig. 1

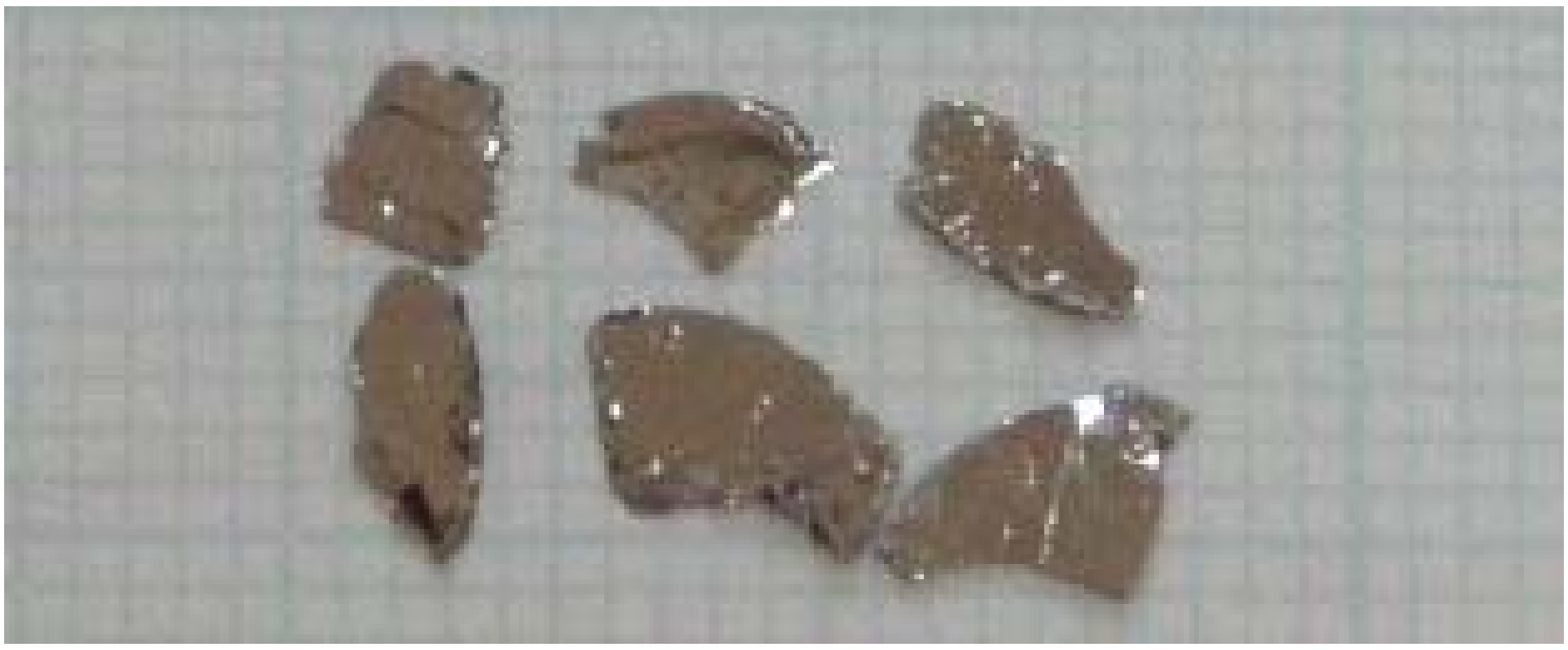

Fig. 2

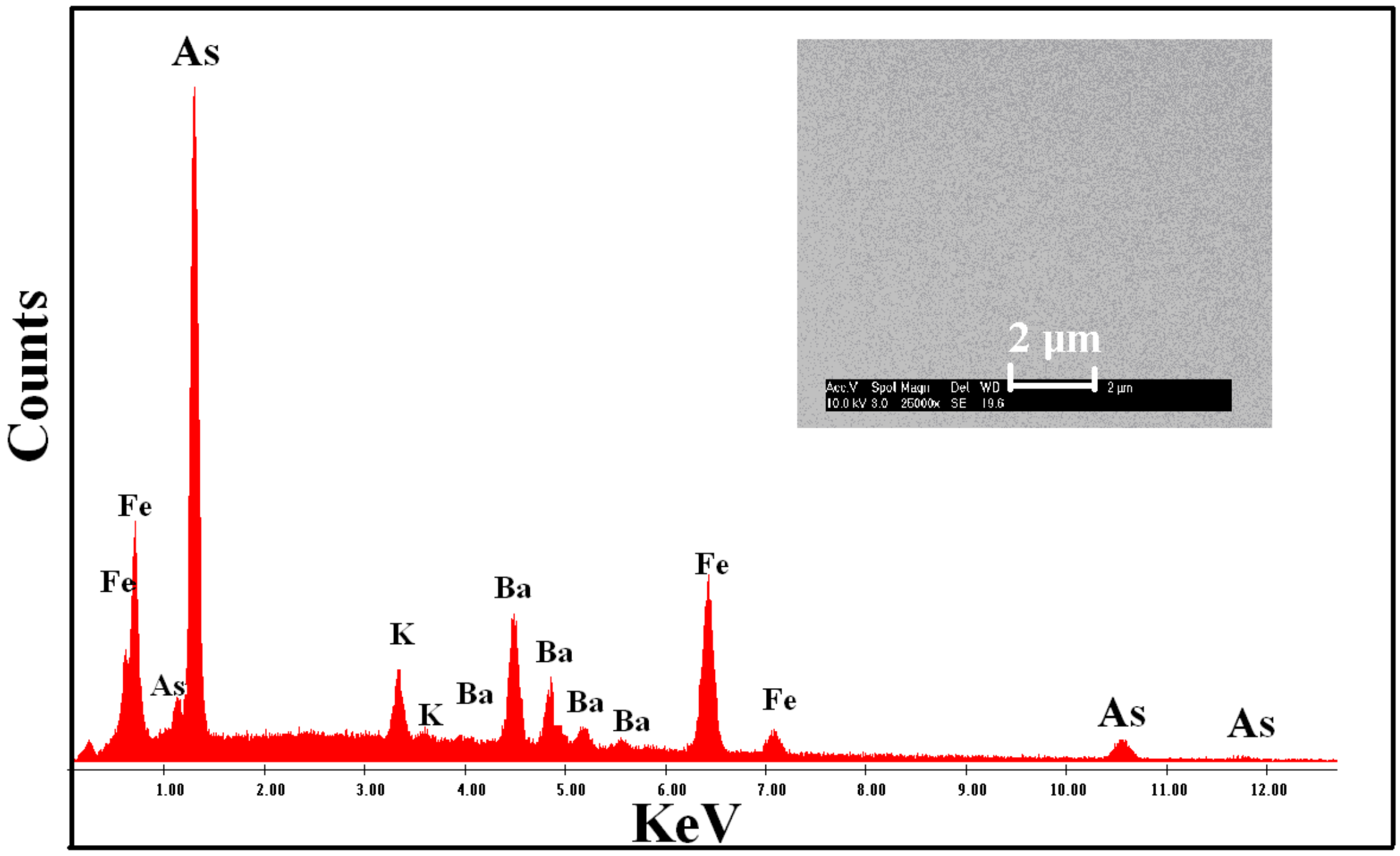

Fig. 3

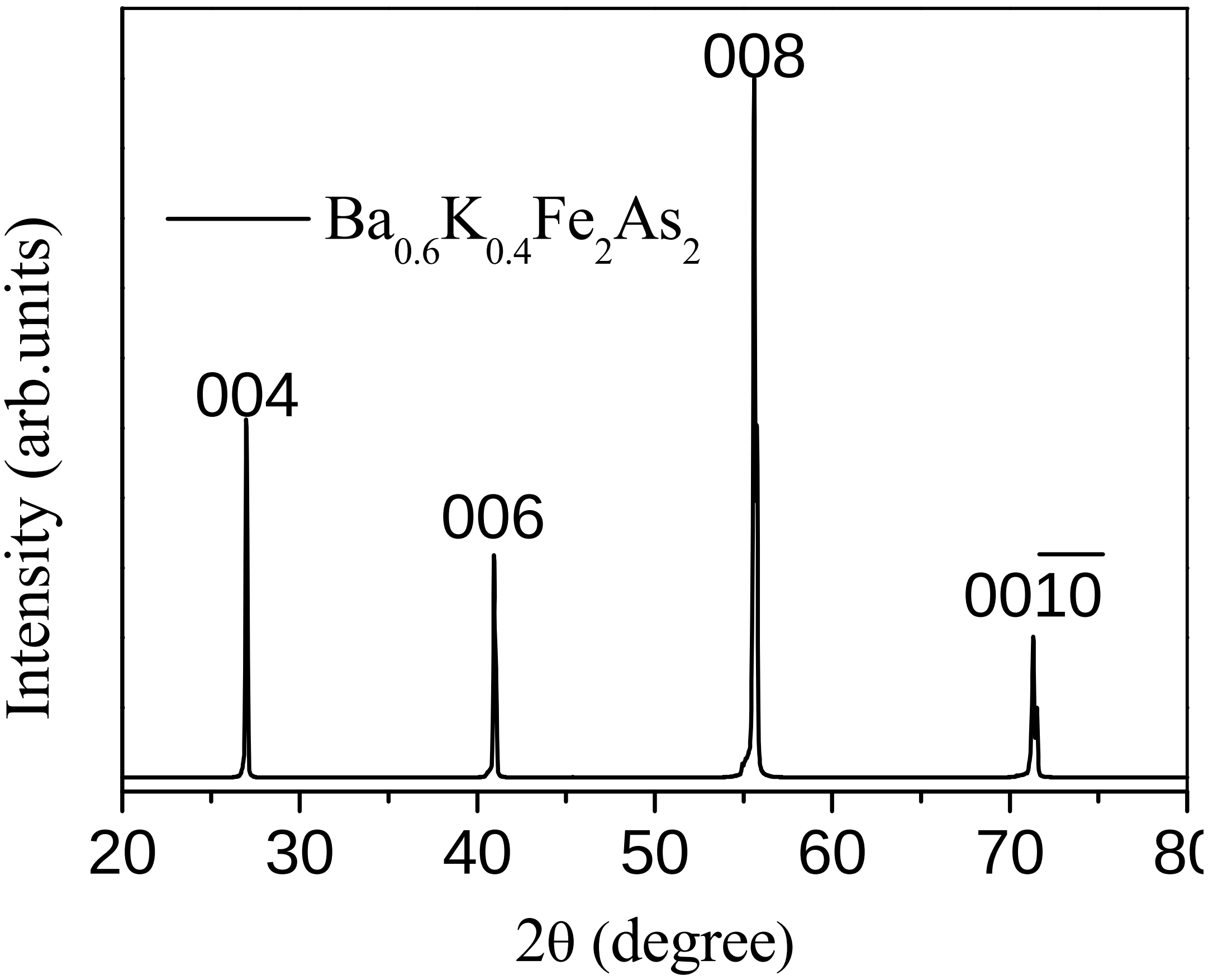

Fig. 4

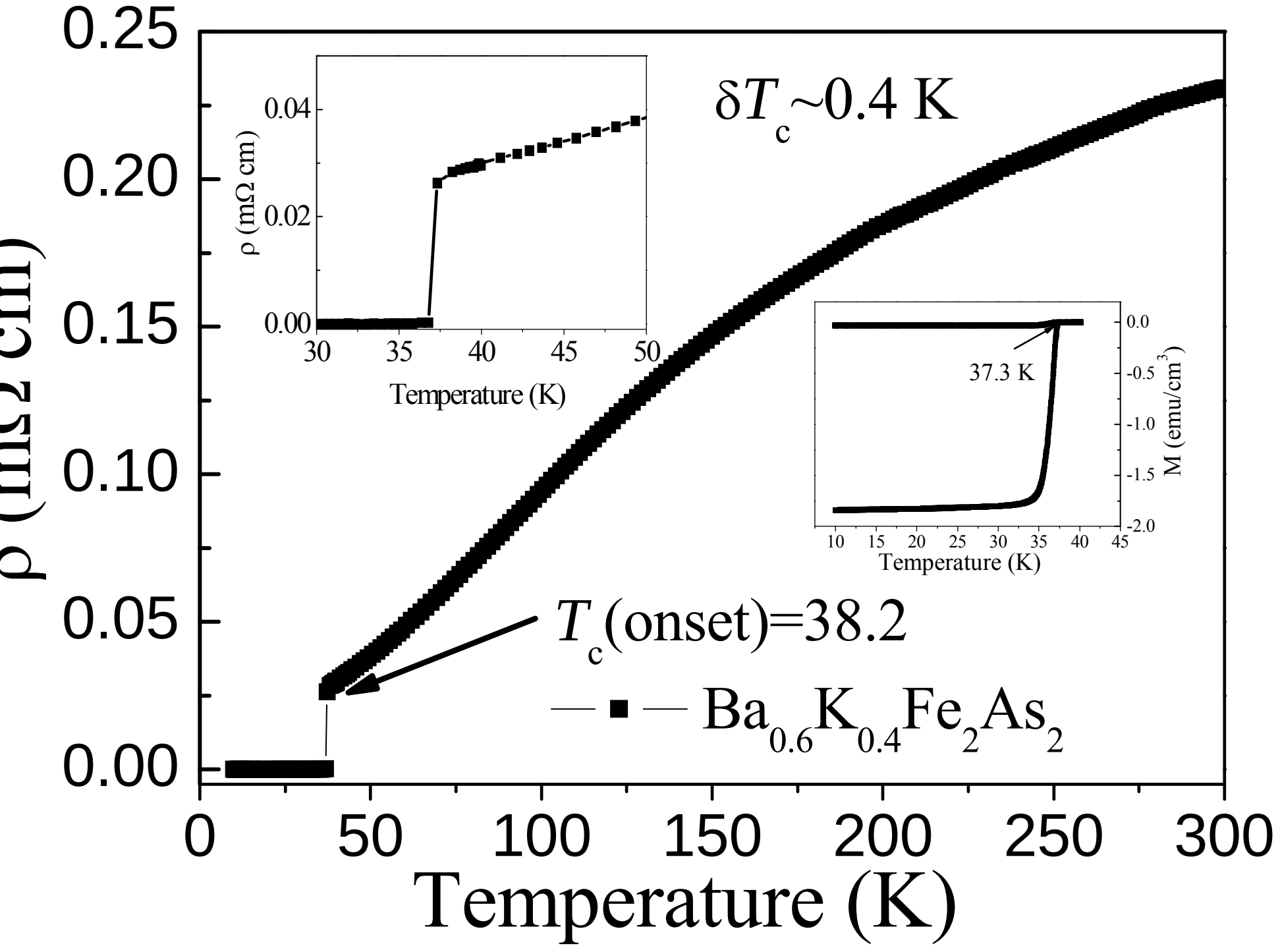

Fig. 5

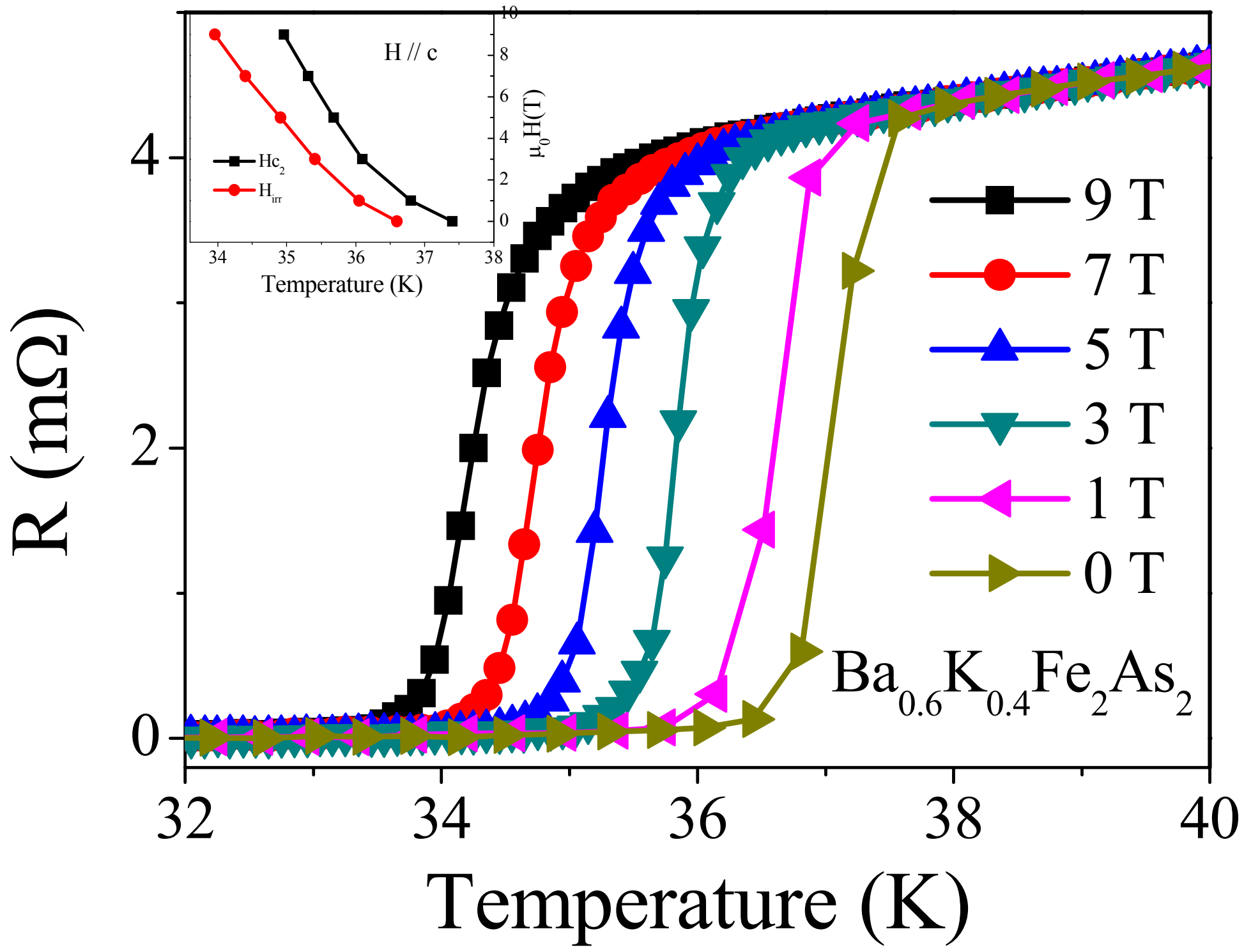

Fig. 6

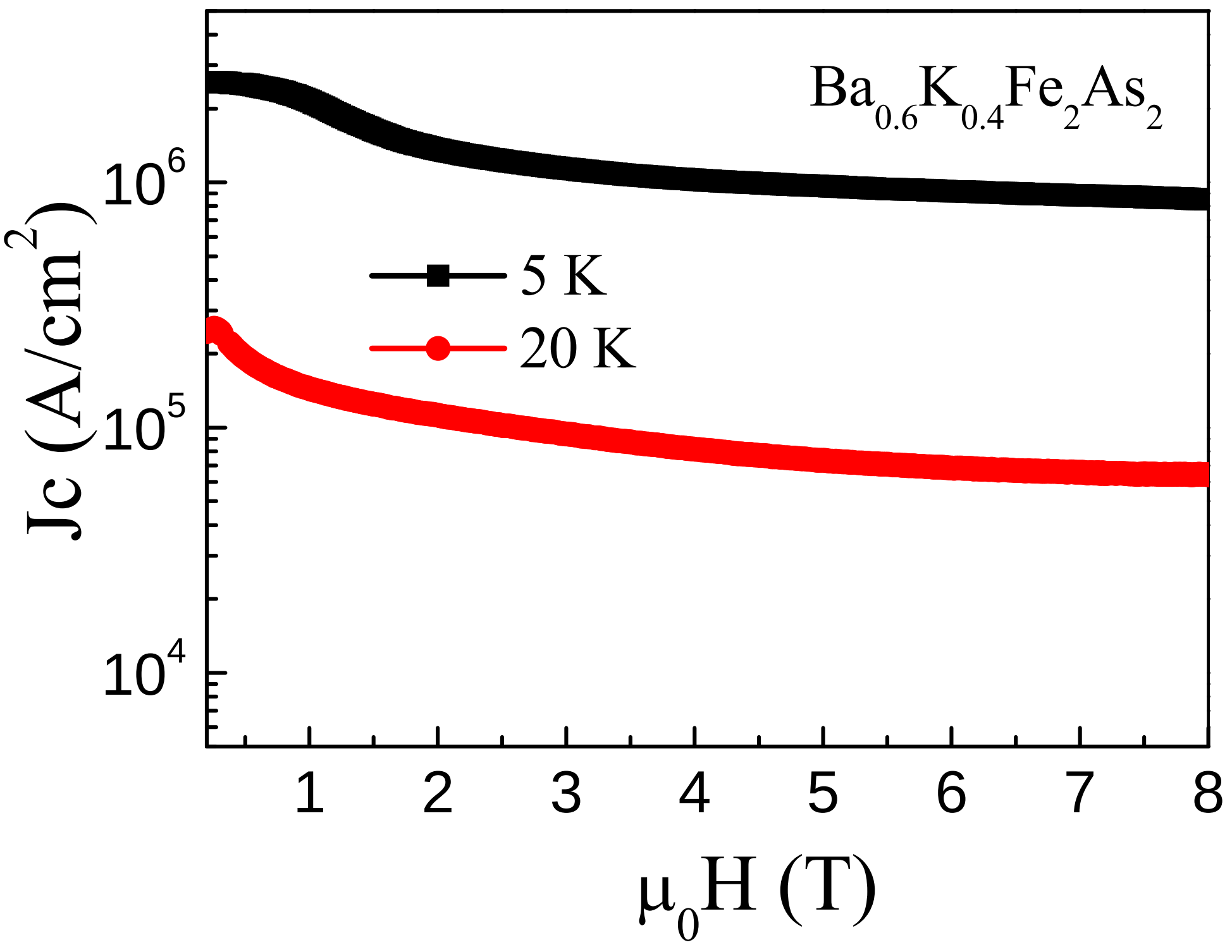

Fig. 7